\documentclass[12pt]{article}

\usepackage{amssymb}
\usepackage{amsmath}
\usepackage{amssymb}
\begin{document}
\begin{center}
{\Large\bf \boldmath  Fundamental Principles of Theoretical Physics and Concepts of Quasiaverages, Quantum 
Protectorate and Emergence}
\footnote{Invited talk, presented at 
XLVIII All-Russia Conference on Problems in 
Particle Physics, Plasma Physics, Condensed Matter, and Optoelectronics,  
which was held in Moscow 15 - 18 May 2012, Russia. 
The Conference was dedicated to the 100th anniversary of Professor Yakov Petrovich Terletsky (1912 - 1993). } \\
 
\vspace*{6mm}
{A. L. Kuzemsky   }\\      
{\small   Bogoliubov Laboratory of Theoretical Physics,}\\
{\small    Joint Institute for Nuclear Research, Dubna, Russia.\\
kuzemsky@theor.jinr.ru; \,  http://theor.jinr.ru/\symbol{126}kuzemsky}         
\end{center}

\vspace*{2mm}

\begin{abstract}
In the present paper we discuss  the interrelation of the advanced interdisciplinary concepts of  modern physics such as 
symmetry breaking, quantum protectorate, emergence and the Bogoliubov's concept 
of quasiaverages    in the context of modern theoretical physics, and, in particular, 
quantum and statistical physics.
The main aim of this analysis was to demonstrate the connection and interrelation of these
conceptual advances of the many-particle physics and to try to show explicitly that those
concepts, though different in details, have  certain common features. Some problems in
the field of statistical physics of complex materials and systems e.g.  
foundation of the microscopic theory of magnetism   and superconductivity   were pointed in relation to these ideas.
The main suggestion
is that  the emphasis of symmetry breaking concept is on the symmetry itself, whereas the method  of quasiaverages 
emphasizes the degeneracy of a system. The concept of quantum protectorate reveals  essential
difference in the behavior of the complex many-body systems at the low-energy
and high-energy scales.  Thus the  notion of quantum protectorate might provide
distinctive signatures and good criteria for a hierarchy of energy scales and  the appropriate emergent behavior.
%
%

\noindent
\textbf{Key words}: Theoretical physics, quantum physics, quantum statistical mechanics, symmetry, broken symmetry, Bogoliubov's quasiaverages,  
 quantum protectorate, emergence,  quantum theory of magnetism, microscopic theory of superconductivity. 
\end{abstract}
%
%
\section{Introduction}
%
The development of experimental techniques over the last decades opened the possibility
for studies and investigations of the wide class of extremely complicated and multidisciplinary
problems in physics, astrophysics, biology, material science, etc. In this regard
theoretical physics is a kind of science which forms and elaborates the appropriate
language for treating these problems on the firm ground~\cite{ref1}. This idea was expressed in the
statement of F. Wilczek~\cite{ref2}: "primary goal of fundamental physics is to discover profound
concepts that illuminate our understanding of nature". For example, the theory of
symmetry is a basic tool for understanding and formulating the fundamental notions of
physics~\cite{wign54,bas90,ref3,leder,ryd06,rose,baro,ref4,ref5}. 
It is well known that symmetry principles play a crucial role in physics. 
Many fundamental laws of physics in addition to their detailed features possess
various symmetry properties~\cite{leder,ryd06}. These symmetry properties lead to certain constraints and
regularities on the possible properties of matter. 
"Symmetry pervades the inner world of the structure of matter, the outer world of the cosmos, and
the abstract world of mathematics itself. The basic laws of physics, the
most fundamental statements we can make about nature, are founded upon symmetry"~\cite{leder}.
Thus the principles of symmetry belongs
to the underlying principles of physics. Moreover, the idea of symmetry is a useful and
workable tool for many areas of quantum field theory~\cite{ryd06}, physics of elementary particles~\cite{leder,ref4,ref5},
statistical physics and condensed matter physics~\cite{sene,jos91,pwa84,kuz10,kuz10a}. Symmetry considerations show that
symmetry arguments are very powerful tool for bringing order into the very complicated
picture of the real world.\\
It is known that when the Hamiltonian of a system is invariant under a symmetry operation, but the
ground state is not, the symmetry of the system can be spontaneously broken~\cite{fstr05}. 
Symmetry breaking is termed \emph{spontaneous} when there is no explicit term in a Lagrangian which
manifestly breaks the symmetry. 
Symmetries and  breaking of symmetries play an important role in  statistical physics, quantum field theory, physics
of elementary particles, etc.~\cite{namb07,namb09,namb}\\
In physics, spontaneous symmetry breaking occurs when a system that is symmetric with respect to some symmetry 
group goes into a vacuum state that is not symmetric. When that happens, the system no longer appears to behave in 
a symmetric manner. It is a phenomenon that naturally occurs in many situations.
The mechanism of spontaneous symmetry breaking is usually understood as the mechanism
responsible for the occurrence of asymmetric states in quantum systems in the thermodynamic limit and is used
in various field of quantum physics~\cite{fstr05}.  
The intriguing mechanism of spontaneous symmetry breaking is a unifying concept that lie at the basis of most 
of the recent developments in theoretical physics, from statistical mechanics to many-body theory and to elementary 
particles theory~\cite{namb07,namb09,namb,huang07,scha08}.\\ 
It should be stressed that symmetry implies degeneracy. The greater the symmetry, the greater the degeneracy.
The study of the degeneracy of the energy levels plays a very important role in quantum physics.
There is an important aspect
of the degeneracy problem in quantum mechanics when  a system possess more subtle symmetries. This is the case when
degeneracy of the levels arises from the invariance of the Hamiltonian $H$ under groups involving simultaneous
transformation of coordinates and momenta that contain as subgroups the usual geometrical groups based on point
transformations of the coordinates. For these groups the free part of $H$ is not invariant, so that the symmetry is established only for
interacting systems. For this reason they are usually called dynamical groups. \\
It is of importance to emphasize that when spontaneous symmetry breaking takes place, the ground state of the system
is degenerate. 
Substantial progress in the understanding of the  broken symmetry concept was connected with
Bogoliubov's fundamental   ideas on quasiaverages~\cite{nnb60,nnb61,nnbj,bs75,nnb61w,bogo47,bb82,nnb58,bts58,petr95,kuz09}. 
Studies of degenerate systems led Bogoliubov in 1960-61 to the formulation of \textbf{the method of quasiaverages}.
This method has proved to be a universal tool for systems whose ground states become unstable under small
perturbations. 
Thus the role of symmetry (and the breaking of symmetries) in combination 
with the degeneracy of the system was reanalyzed and essentially clarified by N. N. Bogoliubov in 1960-1961. He 
invented and formulated
a powerful innovative idea of \emph{quasiaverages} in statistical mechanics~\cite{nnb60,nnb61,nnbj,bs75,bb82,petr95}.
The very elegant work of N. N. Bogoliubov~\cite{nnb61} has been
of great importance for a deeper understanding of phase  transitions, superfluidity  and 
superconductivity~\cite{nnb60,nnb61,nnbj,bs75,nnb61w,bogo47,bb82,petr95,nnb58,bts58,bsan,dsh09}, quantum theory of magnetism~\cite{kuz09} and other fields of equilibrium and nonequilibrium
statistical mechanics~\cite{kuz10,kuz10a,nnb60,nnb61,nnbj,bs75,nnb61w,bogo47,bb82,nnb58,bts58,petr95,kuz09,zub71,kuz07}. 
The concept of quasiaverages is indirectly related to the theory of phase transition. The instability of thermodynamic 
averages with respect to perturbations of the Hamiltonian by a breaking of the invariance with respect to a certain 
group of transformations means that in the system transition to an extremal state occurs.
The mathematical apparatus of the method of quasiaverages includes the Bogoliubov theorem~\cite{kuz10,nnb61,bs75,bb82,petr95} on 
singularities of type $1/q^{2}$  and the Bogoliubov 
inequality for Green and correlation functions as a direct consequence of the method. It includes algorithms 
for establishing non-trivial estimates for equilibrium quasiaverages, enabling one to study the problem of ordering in statistical systems 
and to elucidate the structure of the energy spectrum of the underlying excited states. 
Thus the  Bogoliubov's idea  of \emph{quasiaverages} is an essential
conceptual advance of modern physics. \\  
It is well known that there are many branches of physics and chemistry where phenomena
occur which cannot be described in the framework of interactions amongst a few
particles. As a rule, these phenomena arise essentially from the cooperative behavior of a
large number of particles. Such many-body problems are of great interest not only
because of the nature of phenomena themselves, but also because of the intrinsic difficulties
in solving problems which involve interactions of many particles ( in terms of known
P.W. Anderson's statement: "more is different"~\cite{and72}). It is often difficult to formulate a fully
consistent and adequate microscopic theory of complex cooperative phenomena. 
Statistical mechanics relates the behavior of macroscopic
objects to the dynamics of their constituent microscopic entities. Primary examples include the
entropy increasing evolution of nonequilibrium systems and phase transitions in equilibrium systems.
Many aspects of these phenomena can be captured
in greatly simplified models of the microscopic world.
They emerge as collective properties of large aggregates, i.e. macroscopic systems, which are independent
of many details of the microscopic dynamics.
More recently it has been possible to make a step forward in solving of these problems. This step
leads to a deeper understanding of the relations between microscopic dynamics and
macroscopic behavior on the basis of \textbf{emergence concept}~\cite{and72,ander,life02,wen05,lic05,mina06,aziz09,vol08,emmat09,hu12}.\\
It was shown~\cite{hu12} that emergence phenomena in physics can be understood better in connection with other disciplines.
In particular, since emergence is the overriding issue receiving increasing attention
in physics and beyond, it is of big value for philosophy also.
Different scientific disciplines underlie the different senses of emergence.
There are at least   three senses of emergence and a suggestive view on the emergence of time and 
the direction of time have been discussed intensely. 
The important aspect of emergence concept is different manifestations at different levels of structures, hierarchical 
in form, and corresponding interactions.
It is not easy task to formulate precisely observations pertaining to the concepts, methodology and mechanisms required 
to understand emergence and describe a platform for its investigation~\cite{hu12}.\\
The "quantum protectorate" concept was formulated
in the paper~\cite{pnas}. Its authors, R. Laughlin and
D. Pines, discussed the most fundamental principles of
matter description in the widest sense of this word.
The notion of quantum protectorate~\cite{pnas,laf05,laf08}
complements the concepts of broken symmetry and quasiaverages by making
emphasis on the hierarchy of the energy scales of  many-particle systems~\cite{kuz10,kuz02}.\\
The chief purpose  of this paper was to formulate the connection and interrelation of the complementary conceptual advances 
(or "profound concepts")
of the many-body physics, namely the quasiaverages, emergence and quantum protectorate, 
and to try to show explicitly that
those concepts, though different in details, have a certain common features.
%
\section{Bogoliubov's  Quasiaverages}
%
In the work of N. N. Bogoliubov
"Quasiaverages in Problems of Statistical Mechanics" the innovative notion of \emph{quasiaverage}~\cite{nnb61}
was introduced and applied to various problem of statistical physics. In particular,
quasiaverages of Green's functions constructed from ordinary averages, degeneration 
of statistical equilibrium states, principle of weakened correlations, and particle pair states were considered. 
In this framework the $1/q^{2}$-type properties in the theory of the superfluidity of Bose and Fermi systems, the properties of basic 
Green functions for a Bose system in the presence of condensate, and a model with separated condensate 
were analyzed~\cite{nnbcp} .\\
The method of quasiaverages is a constructive workable scheme for studying systems with spontaneous symmetry 
breakdown such as superluidity and superconductivity~\cite{kuz10,kuz10a,kuz09,nnbcp,tolm69,nnbza,zag01,san02,yuk06,yukle06,bsan10,san10,verb11,yuk12}. 
A quasiaverage is a thermodynamic (in statistical mechanics) or vacuum (in quantum field theory) average of dynamical 
quantities in a specially modified averaging procedure, enabling one to take into account the effects of the influence of 
state degeneracy of the system.
The method gives the so-called macro-objectivation of the degeneracy in the domain of quantum statistical mechanics and in quantum physics.
In statistical mechanics, under spontaneous symmetry breakdown one can, by using the method of quasiaverages, describe 
macroscopic observable within the framework of the microscopic approach.\\
In considering problems of findings the eigenfunctions in quantum mechanics it is well known that the theory of 
perturbations should be modified substantially for the degenerate systems. In the problems of statistical mechanics
we have always the degenerate case due to existence of the additive conservation laws.
The traditional approach to quantum statistical mechanics~\cite{petr95} is based on the unique canonical
quantization of classical Hamiltonians for systems with finitely many degrees of freedom together with the
ensemble averaging in terms of traces involving a statistical operator $\rho$.
For an operator $\mathcal{A}$ corresponding to some physical quantity $A$ the average value of $A$ will be given as
\begin{equation}\label{q1}
\langle A \rangle_{H}  =   \textrm{Tr} \rho A ; \quad  \rho  = \exp ^{- \beta H} / \textrm{Tr} \exp ^{- \beta H},         
\end{equation}
where $H$ is the Hamiltonian of the system, $\beta = 1/kT$ is the reciprocal of the temperature.\\
The core of the problem lies in establishing the existence of a thermodynamic limit (such as $N/V = $ const, 
$V \rightarrow \infty$, $N$ = number of degrees of freedom, $V$ = volume) and its evaluation for the quantities 
of interest. 
Thus in the statistical mechanics the average $\langle A \rangle$ of any dynamical quantity $A$
is defined in a single-valued way. 
In the situations with degeneracy  the specific problems appear. In quantum mechanics, if two linearly independent
state vectors (wave functions in the Schrodinger picture) have the same energy, there is a degeneracy.
In this case more than one independent state of the system corresponds to a single energy level.  
If the statistical equilibrium state of the system 
possesses lower symmetry than the Hamiltonian of the system (i.e. the situation with the spontaneous symmetry breakdown), 
then it is necessary to supplement the averaging procedure  (\ref{q1}) by a rule forbidding  irrelevant   averaging over the values of macroscopic quantities considered 
for which a change is not accompanied by a change in energy. 
This is achieved by introducing quasiaverages, that is, averages over the Hamiltonian $H_{\nu \vec{e}}$ 
supplemented by infinitesimally-small 
terms that violate the additive conservations laws 
$H_{\nu \vec{e}} = H + \nu (\vec{e}\cdot \vec{M})$, ($\nu \rightarrow 0$). Thermodynamic 
averaging may turn out to be unstable with respect to such a change of the original 
Hamiltonian, which is another indication of degeneracy of the equilibrium state. 
According to Bogoliubov~\cite{nnb61}, the quasiaverage of a dynamical quantity  $A$ for the system with the
Hamiltonian $H_{\nu \vec{e}}$  
is defined as the limit
\begin{equation}
\label{q9}
\curlyeqprec A \curlyeqsucc = \lim_{\nu \rightarrow 0} \langle A \rangle_{\nu \vec{e}},
\end{equation}
where $\langle A \rangle_{\nu \vec{e}}$  denotes the ordinary average taken over the Hamiltonian $H_{\nu \vec{e}}$, containing the small symmetry-breaking terms introduced by the inclusion 
parameter $\nu$, which vanish as $\nu \rightarrow 0$ after passage to the thermodynamic limit $V \rightarrow \infty$. 
Thus the existence of degeneracy is reflected directly in the quasiaverages by their dependence upon the arbitrary
unit vector $\vec{e}$.
It is also clear that
\begin{equation}
\label{q10}
\langle A \rangle = \int \curlyeqprec A \curlyeqsucc d \vec{e}.
\end{equation}
According to definition (\ref{q10}), the ordinary thermodynamic average is obtained by extra averaging of the 
quasiaverage over the symmetry-breaking group. Thus to describe the case of a degenerate state of statistical
equilibrium quasiaverages are more convenient, more physical, than ordinary averages~\cite{petr95}. The latter are the same 
quasiaverages only averaged over all the directions $\vec{e}$.\\
It is necessary to stress,
that the starting point for Bogoliubov's work~\cite{nnb61} was
an investigation of additive conservation laws and
selection rules, continuing and developing the  approach by P. Curie for derivation of
selection rules for physical effects. Bogoliubov demonstrated
that in the cases when the state of statistical
equilibrium is degenerate, as in the case of the Heisenberg ferromagnet,
one can remove the degeneracy of equilibrium
states with respect to the group of spin rotations by
including in the Hamiltonian $H$ an additional noninvariant
term $\nu M_{z} V$ with an infinitely small $\nu$.  For the Heisenberg ferromagnet the ordinary averages must be invariant
with regard to the spin rotation group. The corresponding quasiaverages possess only the property of covariance.
Thus the quasiaverages do not follow the same selection rules as those which govern ordinary averages, due to their 
invariance with regard to the spin rotation group. It is clear that  the unit vector $\vec{e}$, i.e., the
direction of the magnetization $\vec{M}$ vector, characterizes the degeneracy of the considered state of statistical
equilibrium. In order to remove the degeneracy one should fix the direction  of the unit vector $\vec{e}$. It can be
chosen to be along the $z$ direction. Then all the quasiaverages will be the definite numbers. This is the kind 
that one usually deals with in the theory of ferromagnetism.\\
The question of symmetry breaking within the
localized and band models of antiferromagnets was
studied by the author of this report in Refs.~\cite{kuz10,kuz09,kuz02,rnc02}. It has been found there that the concept of
spontaneous symmetry breaking in the band model of
magnetism is much more complicated than in the
localized model~\cite{kuz09}. In the framework of the band model of
magnetism one has to additionally consider the so called
anomalous propagators of the form~\cite{kuz10,kuz09,kuz02,rnc02}
\begin{eqnarray} \textrm{FM}: G_{fm} \sim \langle \langle a_{k\sigma};a^{\dag}_{k-\sigma} \rangle \rangle,  \nonumber \\
\nonumber \textrm{AFM}: G_{afm} \sim \langle \langle a_{k+Q\sigma};a^{\dag}_{k+Q'\sigma'} \rangle \rangle. \nonumber
\end{eqnarray}
In the case of the band antiferromagnet the ground
state of the system corresponds to a spin-density wave
(SDW), where a particle scattered on the internal inhomogeneous
periodic field gains the momentum $Q - Q'$
and changes its spin: $\sigma \rightarrow \sigma'$. The long-range order
parameters are defined as follows:
\begin{eqnarray} \label{e180}
\textrm{FM}: m =
1/N\sum_{k\sigma} \langle a^{\dag}_{k\sigma}a_{k-\sigma} \rangle,\\  \textrm{AFM}: M_{Q}
= \sum_{k\sigma} \langle a^{\dag}_{k\sigma}a_{k+Q-\sigma} \rangle. \label{e181}  \end{eqnarray}
It is important to stress, that the long-range order
parameters here are functionals of the internal field, which
in turn is a function of the order parameter. Thus, in the
cases of  Hamiltonians of
band ferro- and antiferromagnetics one has to add the following
infinitesimal sources removing the degeneracy:
\begin{eqnarray} \label{e182}
\textrm{FM}:  \nu\mu_{B}
H_{x}\sum_{k\sigma}a^{\dag}_{k\sigma}a_{k-\sigma},\\   \textrm{AFM}:
\nu \mu_{B} H \sum_{kQ} a^{\dag}_{k\sigma}a_{k+Q-\sigma}.
\label{e183}
\end{eqnarray}
Here, $\nu \rightarrow 0$ after the usual in statistical mechanics
infinite-volume limit $V \rightarrow \infty$. The ground state in the
form of a spin-density wave was obtained for the first
time by Overhauser. There, the vector  $\vec{Q}$ is a measure of inhomogeneity
or translation symmetry breaking in the system. 
The analysis performed  by various authors  showed  that the antiferromagnetic
and more complicated states (for instance, ferrimagnetic)
can be described in the framework of a generalized
mean-field approximation. In doing that we have to
take into account both the normal averages $\langle a^{\dag}_{i\sigma}a_{i\sigma}\rangle$
and the anomalous averages $\langle a^{\dag}_{i\sigma}a_{i-\sigma}\rangle$.  It is clear that
the anomalous terms  break the original
rotational symmetry of the Hubbard Hamiltonian.
Thus, the generalized mean-field's approximation has
the following form 
$n_{i-\sigma}a_{i\sigma} \simeq \langle n_{i-\sigma}\rangle a_{i\sigma} -
\langle a^{\dag}_{i-\sigma}a_{i\sigma}\rangle a_{i-\sigma}.$  
A self-consistent theory of band antiferromagnetism
was developed by the author of this report
using the method of the irreducible Green functions~\cite{kuz10,kuz09,kuz02,rnc02}. The
following definition of the irreducible Green functions  was used:
\begin{eqnarray} \label{e184}
^{ir}\langle \langle a_{k+p\sigma}a^{\dag}_{p+q-\sigma}a_{q-\sigma} \vert
a^{\dag}_{k\sigma} \rangle \rangle_ {\omega} =
\langle \langle a_{k+p\sigma}a^{\dag}_{p+q-\sigma}a_{q-\sigma}\vert
a^{\dag}_{k\sigma} \rangle \rangle_{\omega} - \nonumber\\ \delta_{p,
0}\langle n_{q-\sigma}\rangle G_{k\sigma} - \langle a_{k+p\sigma}a^{\dag}_{p+q-\sigma}\rangle
\langle \langle a_{q-\sigma} \vert a^{\dag}_{k\sigma} \rangle \rangle_{\omega}. \end{eqnarray} 
The algebra of relevant operators must be chosen as follows
($(a_{i\sigma}$,
$a^{\dag}_{i\sigma}$, $n_{i\sigma}$, $a^{\dag}_{i\sigma}a_{i-\sigma})$.
The corresponding
initial GF will have the following matrix structure
$$\mathcal{G}_{AFM} =  \begin{pmatrix}
\langle \langle a_{i\sigma}\vert a^{\dag}_{j\sigma}\rangle \rangle & \langle \langle a_{i\sigma}\vert
a^{\dag}_{j-\sigma}\rangle \rangle \cr
\langle \langle a_{i-\sigma}\vert a^{\dag}_{j\sigma}\rangle \rangle & \langle \langle a_{i-\sigma}\vert
a^{\dag}_{j-\sigma} \rangle \rangle \cr   \end{pmatrix}.$$
The off-diagonal terms select the vacuum state of
the band's antiferromagnet in the form of a spin-density
wave. It is necessary to stress that the problem of the
band's antiferromagnetism   is quite involved,
and the construction of a consistent microscopic theory
of this phenomenon remains a topical problem.\\
D.N. Zubarev showed~\cite{zub71}  that the concepts of symmetry
breaking perturbations and quasiaverages play an
important role in the theory of irreversible processes as
well. The method of the construction of the nonequilibrium
statistical operator~\cite{kuz09,zub71,kuz07} becomes especially
deep and transparent when it is applied in the framework
of the quasiaverage concept. The main idea of
this approach was to consider infinitesimally
small sources breaking the time-reversal symmetry of
the Liouville equation,
which become vanishingly small after a thermodynamic
limiting transition.\\ 
To summarize, the Bogoliubov's method of quasiaverages   gives the deep foundation and clarification of the
concept of broken symmetry. It makes the emphasis on the notion of a degeneracy and
plays an important role in equilibrium statistical mechanics of many-particle systems. According to
that concept, infinitely small perturbations can trigger
macroscopic responses in the system if they break some
symmetry and remove the related degeneracy (or quasidegeneracy)
of the equilibrium state. As a result, they
can produce macroscopic effects even when the perturbation
magnitude is tend to zero, provided that happens
after passing to the thermodynamic limit.  
Therefore the method of quasiaverages plays a fundamental role in equilibrium and
nonequilibrium statistical mechanics and is one of the pillars of modern physics.
%
%
%
\section{Emergent Phenomena}
%
%
%
Emergence and complexity refer to the appearance of higher-level properties and behaviors of a system that obviously 
comes from the collective dynamics of that system's 
components~\cite{and72,ander,life02,wen05,lic05,mina06,aziz09,vol08,emmat09,hu12,pnas,laf05,laf08}. These 
properties are not directly deducible from the lower-level motion of that system. Emergent properties are properties 
of the ''whole'' that are not possessed by any of the individual parts making up that whole. Such phenomena 
exist in various domains and can be described, using complexity concepts and thematic knowledges. 
Thus this problematic   is highly pluridisciplinary.  
Emergence - macro-level effect from micro-level causes - is an important and profound interdisciplinary notion of modern 
science~\cite{and72,ander,life02,wen05,lic05,mina06,aziz09,vol08,emmat09,hu12,pnas,laf05,laf08}.\\
Emergence is a   key notion when discussing various aspects of
what are  termed self-organizing systems, spontaneous orders, 
chaotic systems, system  complexity, and so on. This   variety of the problems reflects
the multidisciplinary nature of the emergence concept,
because the concept has appeared relatively independently in various contexts
within philosophy~\cite{hu12}, the social and the natural sciences~\cite{lic05,mina06}.
Emergence unites these disciplines in the sense that it emphasizes their common focus on phenomena where orders arises
from elements within a system acting independently from one another. In addition, emergence stresses that  such an action
is realized within a framework of procedural rules or laws that generate positive and negative feedback such
that independent behavior takes the actions of others into consideration without intending
to do so. Moreover,  the impact of that behavior tends to facilitate more complex
relationships of mutual assistance than could ever be deliberately created. Such systems
may generate order \emph{spontaneously}.  In doing so they can act in unanticipated ways because
there is no overarching goal, necessity, or plan that orders the actions of their
components or the responses they make to feedback generated within the system.\\
Indeed, self-organization, fractals, chaos, and many other interesting dynamical phenomena can be understood better with the help
of the emergence concept~\cite{life02,lic05,mina06}.
For example, a system with positive and negative feedback loops is modeled with nonlinear equations. Self-organization may occur 
when feedback loops exist among component parts and between the parts and the structures that emerge at higher 
hierarchical levels.  In chemistry, when an enzyme catalyzes reactions that encourage 
the production of more of itself, it is called auto-catalysis. It was suggested that auto-catalysis played an important 
role in the origins of life~\cite{life02}.
Thus the essence of self-organization lies in the connections, interactions, and feedback loops between the parts 
of the system~\cite{life02,lic05,mina06}. It is clear then that system must have a large number of parts. Cells, living tissue, the immune system, 
brains, populations,   communities, economies, and climates all contain huge number of parts. These parts are often 
called agents because they have the basic properties of information transfer, storage and processing. 
An agent could be a ferromagnetic particle in a spin glass, a neuron in a brain, or a firm in an economy. Models that 
assign agency at this level are known as individual-based models. 
It is possible to say that emergence is  a kind of observation, when the observer's attention shifts from the 
micro-level of the agents to the macro-level of the system. Emergence fits well into hierarchy theory as a way of 
describing how each hierarchical level in a system can follow discrete rule sets.\\
Emergence also points to the multiscale interactions~\cite{life02,lic05,mina06,pnas,laf05,laf08} and effects in self-organized systems. The small-scale interactions produce large-scale 
structures, which then modify the activity at the small scales. For instance, specific chemicals and neurons in the immune system can create 
organism-wide bodily sensations which might then have a huge effect on the chemicals and neurons. Some authors has argued 
that macro-scale emergent order is a way for a system to dissipate micro-scale entropy creation caused by energy flux, 
but this is still  a hypothesis which must be verified.\\
Thus  emergent entities (properties or substances) \emph{arise}  out of more 
fundamental entities and yet are \emph{novel}  or  \emph{irreducible}  with respect to them.  
Each of these terms are uncertain in its own right, and their specifications 
yield the varied notions of emergence that have been  discussed 
in literature~\cite{and72,ander,life02,wen05,lic05,mina06,aziz09,vol08,emmat09,hu12,pnas,laf05,laf08}. There has been 
renewed interest in emergence within discussions of the behavior of complex systems~\cite{lic05,mina06,aziz09} and debates over the 
reconcilability of mental causation, intentionality, 
or consciousness with physicalism. This concept is also at the heart of the numerous discussions on the interrelation of the
reductionism and functionalism~\cite{and72,ander,life02,hu12,pnas,laf05,laf08}.\\
A vast amount of current researches focuses on the search for the organizing principles responsible for emergent behavior 
in matter~\cite{pnas,pines}, with particular attention to correlated matter, the study of materials in which unexpectedly 
new classes of behavior emerge in response to the strong and competing interactions among their elementary constituents.
As it was   formulated at Ref.~\cite{pines}, ''we call \emph{emergent behavior} \ldots the phenomena that owe their
existence to interactions between many subunits, but whose existence cannot be deduced from a detailed knowledge
of those  subunits alone''.\\ 
Models and simulations of collective behaviors are often based on considering them as interactive particle 
systems~\cite{mina06}. The 
focus is then on behavioral and interaction rules of particles by using approaches based on artificial agents designed to 
reproduce swarm-like behaviors in a virtual world by using symbolic, sub-symbolic and agent-based models. New approaches 
have been considered in the literature~\cite{lic05,mina06,aziz09} based, for instance, on  topological rather than metric distances and 
on fuzzy systems. Recently a new research approach~\cite{mina06} was  proposed  allowing generalization possibly suitable 
for a general theory of emergence.  The coherence of collective behaviors, i.e., their identity detected by the observer, 
as given by meta-structures, properties of meta-elements, i.e., sets of values adopted by mesoscopic state variables 
describing collective, structural aspects of the collective phenomenon under study and related to a higher level of description 
(meta-description) suitable for dealing with coherence, was considered. Mesoscopic state variables were abductively 
identified by the observer detecting emergent properties, such as sets of suitably clustered distances, speed, directions, 
their ratios and ergodic properties of sets. This research approach is under implementation and validation and may be 
considered to model general processes of collective behavior and establish an possible initial basis for a general theory of 
emergence.\\
Statistical physics and condensed matter physics supply us with many examples of emergent phenomena~\cite{kuz10,kuz10a}.
For example, taking a macroscopic approach to the problem, 
and identifying the right degrees of freedom of a many-particle system, the equations of motion of interacting particles 
forming a fluid can be described by the Navier-Stokes equations 
for fluid dynamics from which complex new behaviors arise such as turbulence. This is  the clear example of 
an emergent phenomenon in classical physics.\\
Including quantum mechanics into the consideration leads to even more complicated situation. 
In 1972 P. W. Anderson published his essay "More is Different"
which describes how new concepts, not applicable in ordinary classical or
quantum mechanics, can arise from the consideration of aggregates of large numbers of
particles~\cite{and72}. 
Quantum mechanics is a basis of macrophysics. However 
macroscopic systems have the properties that are radically different from those of their constituent particles. Thus, 
unlike systems of few particles, they exhibit irreversible dynamics, phase transitions and various ordered structures, 
including those characteristic of life~\cite{and72,ander,life02,wen05,lic05,mina06,aziz09,vol08,emmat09,hu12,pnas,laf05,laf08}. These and other macroscopic phenomena signify that complex systems, that is, 
ones consisting of huge numbers of interacting particles, are qualitatively different from the sums of their 
constituent parts~\cite{and72}.\\
Many-particle systems where the interaction 
is strong have often complicated behavior, and require nonperturbative approaches to treat their properties.  
Such  situations  are often arise in condensed matter systems.
Electrical, magnetic and mechanical properties of materials are \emph{emergent collective behaviors}  of the underlying quantum mechanics of 
their electrons and constituent atoms. A principal aim of solid state physics and materials science is to elucidate this emergence. 
A full achievement of this goal would imply the ability to engineer a material that is optimum for any particular 
application. The current understanding of electrons in solids uses simplified but workable picture
 known as the Fermi liquid theory. This theory explains why 
electrons in solids can often be described in a simplified manner which appears to ignore the large repulsive forces 
that electrons are known 
to exert on one another. There is a growing appreciation that this theory probably fails for entire classes of possibly 
useful materials and 
there is the suspicion that the failure has to do with unresolved competition between different possible emergent 
behaviors.\\ 
In Ref.~\cite{wen05}  Levine and Wen proposed to consider photons and electrons as emergent phenomena.
Their arguments are based on recent advances in condensed-matter theory which have revealed that new 
and exotic phases of matter can exist in spin models (or more precisely, local bosonic models) via a simple physical 
mechanism, known as ''\emph{string-net condensation}''.  These new phases of matter have the unusual property 
that their collective excitations are gauge bosons and fermions. In some cases, the collective excitations can behave just like the photons, 
electrons, gluons, and quarks in the relevant vacuum. This suggests that photons, electrons, and other elementary particles may have a 
unified origin-string-net condensation in that  vacuum. In addition, the string-net picture indicates how to make artificial photons,
artificial electrons, and artificial quarks and gluons in condensed-matter systems. 
%
\section{Quantum Mechanics and its Emergent Macrophysics}
%
%
%
The notion of emergence in quantum physics has attracted recently big 
attention~\cite{bla09,gro12,sew02,adler04,adler12,gth07,gth07a,gth12,elze08}.   
Although quantum mechanics (QM) has been the generally accepted basis for most of progress in
fundamental physics during the last 100 years, the extension of the current theoretical
frontier to Planck's scale physics, and recent enlargements of our experimental capabilities, may
make our time the period in which possible limits of quantum theory will be subjected
to a thorough scrutiny. Some authors speculate   that QM may be actually not a complete
ontological system, but in fact it represents a very accurate low-energy approximation to a
deeper level of dynamics (hierarchy of energy scales). But what is exactly the "deeper level dynamics" is not 
clear at all. There is a growing interest in these problems which was partially supported by the belief that  
to make a convincing synthesis of QM and general relativity~\cite{vol08}     new conceptual paradigms should be formulated
 to describe physics at very small space-time scales.\\ 
The interrelation of notion of emergence and QM  was considered by Sewell 
in his book "Quantum Mechanics And Its Emergent Macrophysics"~\cite{sew02}. According to his point of view,
the quantum theory of macroscopic systems is a vast, ever-developing area of science that serves to relate the properties 
of complex physical objects to those of their constituent particles. Its essential challenge is that of finding the 
conceptual structures needed for the description of the various states of organization of many-particle quantum systems. 
In that book,  Sewell proposes a new approach to the subject, based on a "\emph{macrostatistical mechanics}", which 
contrasts sharply with the standard microscopic treatments of many-body problems.\\
According to Sewell,
quantum theory began with Planck's   derivation of the thermodynamics of black body radiation from the hypothesis that 
the action of his oscillator model of matter was quantized in integral multiples of a fundamental constant, $\hbar$. This 
result provided a microscopic theory of a macroscopic phenomenon that was incompatible with the assumption of underlying 
classical laws. In the century following Planck's discovery, it became abundantly clear that quantum theory is essential 
to natural phenomena on both the microscopic and macroscopic scales.\\
As a first step towards contemplating the quantum mechanical basis of macrophysics, Sewell notes the empirical fact that 
macroscopic systems enjoy properties that are radically different from those of their constituent particles. Thus, unlike 
systems of few particles, they exhibit 
irreversible dynamics, phase transitions and various ordered structures, including those characteristic of life. 
These and other macroscopic 
phenomena signify that complex systems, that is, ones consisting of enormous numbers of interacting particles, are 
qualitatively different from the sums of their constituent parts (this point of view was also stressed by 
Anderson~\cite{and72}).\\
Sewell proceeds by presenting the operator algebraic framework for the theory. He then undertakes a macrostatistical treatment 
of both equilibrium and nonequilibrium thermodynamics, which yields a major new characterization of a complete set of 
thermodynamic variables and a nonlinear generalization of the Onsager theory. He focuses especially on 
ordered and chaotic structures that arise in some key areas of condensed matter physics. This includes a general derivation 
of superconductive electrodynamics from the assumptions of off-diagonal long-range order, gauge covariance, and 
thermodynamic stability, which avoids the enormous complications of the microscopic treatments. 
Sewell also re-analyzes a theoretical framework for phase transitions far from thermal equilibrium. 
It gives  a coherent approach to the complicated problem of the emergence of macroscopic phenomena from quantum mechanics 
and clarifies the problem of how macroscopic phenomena can be interpreted from the laws and structures of microphysics.\\ 
Correspondingly, theories of such phenomena must be based not only on the quantum mechanics, but 
also on conceptual structures that serve to represent the characteristic features of 
highly complex systems~\cite{pnas,laf05,laf08}. Among the main   concepts 
involved here are ones representing various types of order, or organization, disorder, or chaos, and different levels 
of macroscopicality. 
Moreover, the particular concepts required to describe the ordered structures of superfluids and laser light are 
represented by macroscopic wave functions   that are strictly quantum mechanical, although radically different from 
the Schrodinger wave functions of microphysics.\\
Thus, according to Sewell, to provide a mathematical framework for the conceptual structures required for 
quantum macrophysics, it is clear that one needs to go beyond 
the traditional form of quantum mechanics, since that does not discriminate qualitatively between microscopic and 
macroscopic systems. 
This may be seen from the fact that the traditional theory serves to represent a system of $N$ particles within the standard 
Hilbert space scheme, which takes the same form regardless of whether $N$ is 'small' or 'large'.\\ 
Sewell's approach to the basic problem of how macrophysics emerges from quantum mechanics is centered on macroscopic 
observables. 
The main objective of his approach is to obtain the properties imposed on them by general demands of quantum theory 
and many-particle statistics. 
This approach resembles in a certain sense the Onsager's irreversible thermodynamics, which
bases also  on macroscopic observables and certain general structures of complex systems~\cite{zub71,kuz07}.\\
The conceptual basis of quantum mechanics which go far beyond its  traditional form was formulated 
by S. L. Adler~\cite{adler04}. According to his view,
quantum mechanics is not a complete theory, but rather is an emergent phenomenon arising from the statistical mechanics of 
matrix models that have a global unitary invariance.
The mathematical presentation  of these ideas is based on dynamical variables that are matrices in complex Hilbert space, 
but many of the ideas carry over to a statistical dynamics of matrix models in real or quaternionic Hilbert space.
Adler starts from a classical dynamics in which the dynamical variables are non-commutative matrices or operators. 
Despite the non-commutativity, a sensible Lagrangian and Hamiltonian dynamics was obtained by forming the Lagrangian and 
Hamiltonian as traces of polynomials in the dynamical variables, and repeatedly using cyclic permutation under the 
trace. It was  assumed that the Lagrangian and Hamiltonian are constructed without use of non-dynamical matrix 
coefficients, so that there is an invariance under simultaneous, identical unitary transformations of all the dynamical 
variables, that is, there is a global unitary invariance. The author supposed that the complicated dynamical equations 
resulting from this system rapidly reach statistical equilibrium, and then shown that with suitable approximations, 
the statistical thermodynamics of the canonical ensemble for this system takes the form of quantum field theory.  
The requirements for the underlying trace dynamics to yield quantum theory at the level of thermodynamics are stringent, 
and include both the generation of a mass hierarchy and the existence of boson-fermion balance.   From the equilibrium 
statistical mechanics of trace dynamics, the rules of quantum mechanics \emph{emerge} as an approximate thermodynamic 
description of the behavior of low energy phenomena. "Low energy" here means small relative to the natural energy scale 
implicit in the canonical ensemble for trace dynamics, which author identify with the Planck scale, and by "equilibrium" 
he means local equilibrium, permitting spatial variations associated with dynamics on the low energy scale. Brownian 
motion corrections to the thermodynamics of trace dynamics then lead to fluctuation corrections to quantum mechanics 
which take the form of stochastic modifications of the Schrodinger equation, that can account in a mathematically precise 
way for state vector reduction with Born rule probabilities~\cite{adler04}.\\
Adler emphasizes~\cite{adler04} that he have not identified a candidate for the specific matrix model that realizes his 
assumptions; there may be only one, which could then provide the underlying unified theory of physical phenomena that 
is the goal of current researches in high-energy physics and cosmology.   
He admits  the possibility also that the underlying dynamics may be discrete, and this could naturally be implemented 
within his framework of basing an underlying dynamics on trace class matrices. 
The ideas of the Adler's book suggest, that one should seek a common origin for both gravitation and quantum field theory 
at the deeper level of physical phenomena from which quantum field theory 
emerges~\cite{adler04}. Recently Adler discussed  his ideas further~\cite{adler12}.
He reviewed the proposal made in his 2004 book~\cite{adler04}, that quantum theory is an emergent theory arising from a 
deeper level of dynamics. The dynamics at this deeper level is taken to be an extension of classical dynamics to 
non-commuting matrix variables, with cyclic 
permutation inside a trace used as the basic calculational tool. With plausible assumptions, quantum theory was shown to 
emerge as the statistical thermodynamics of this underlying theory, with the canonical commutation-anticommutation relations 
derived from a generalized equipartition theorem. Brownian motion corrections to this thermodynamics were argued to lead 
to state vector reduction and to the probabilistic interpretation of quantum theory, making contact with phenomenological 
proposals   for stochastic modifications to Schrodinger dynamics.\\
G. 't Hooft considered various aspects of quantum mechanics in the context of emergence~\cite{gth07,gth07a,gth12}.
According to his view, quantum mechanics is  \emph{emergent}  if a statistical treatment of large scale phenomena
in a locally deterministic theory requires the use of quantum operators.
These quantum operators may allow for symmetry transformations
that are not present in the underlying deterministic system. Such theories
allow for a natural explanation of the existence of gauge equivalence classes
(gauge orbits), including the equivalence classes generated by general coordinate
transformations. Thus, local gauge symmetries and general coordinate
invariance could be emergent symmetries, and this might lead to new alleys
towards understanding the flatness problem of the Universe.
G. 't Hooft demonstrated also that ``For any quantum system there exists at least one deterministic model that reproduces all its 
dynamics after prequantization''. 
H.T. Elze elaborated  an extension~\cite{elze08}  which covers quantum systems that are characterized by a complete set of 
mutually commuting Hermitian operators (``\emph{beables}''). He introduced the symmetry of beables: any complete set of beables is 
as good as any other one which is obtained through a real general linear group transformation. The quantum numbers of a 
specific set are related to symmetry breaking initial and boundary conditions in a deterministic model. 
The Hamiltonian, in particular, can be taken as the emergent beable which provides the 
best resolution of the evolution of the model Universe. 
%
%
%
%
\section{Quantum Protectorate}
%
%
%
R. Laughlin and D. Pines invented
an idea of a quantum protectorate, "a stable state of matter,
whose generic low-energy properties are determined by a
higher-organizing principle and nothing else"~\cite{pnas}. This
idea brings into physics the concept that emphasize   the
crucial role of low-energy and high-energy scales for treating the propertied of the substance. 
It is known that a many-particle  system  (e.g. electron gas) in the low-energy limit can be
characterized by a small set of  \emph{collective}  (or hydrodynamic)
variables and equations of motion corresponding
to these variables. Going beyond the framework
of the low-energy region would require the consideration
of plasmon excitations, effects of electron
shell reconstructing, etc. The existence of two
scales, low-energy and high-energy, in the description
of physical phenomena is used in physics, explicitly or
implicitly.\\
According to R. Laughlin and D. Pines,
''The emergent physical phenomena regulated by higher organizing
principles have a property, namely their insensitivity to
microscopics, that is directly relevant to the broad question of
what is knowable in the deepest sense of the term. The low energy
excitation spectrum of a conventional superconductor,
for example, is completely generic and is characterized by a
handful of parameters that may be determined experimentally
but cannot, in general, be computed from first principles. An
even more trivial example is the low-energy excitation spectrum
of a conventional crystalline insulator, which consists of transverse
and longitudinal sound and nothing else, regardless of
details. It is rather obvious that one does not need to prove the
existence of sound in a solid, for it follows from the existence of
elastic moduli at long length scales, which in turn follows from
the spontaneous breaking of translational and rotational symmetry
characteristic of the crystalline state. Conversely, one
therefore learns little about the atomic structure of a crystalline
solid by measuring its acoustics.
The crystalline state is the simplest known example of a
quantum protectorate, a \emph{stable state of matter whose generic
low-energy properties are determined by a higher organizing
principle and nothing else} \ldots  Other important quantum protectorates
include superfluidity in Bose liquids such as $^{4}He$ and
the newly discovered atomic condensates, superconductivity, band insulation, ferromagnetism, 
antiferromagnetism, and the quantum Hall states. The
low-energy excited quantum states of these systems are particles
in exactly the same sense that the electron in the vacuum of
quantum electrodynamics is a particle  \ldots Yet they are not elementary, and, as in the
case of sound, simply do not exist outside the context of the
stable state of matter in which they live. These quantum protectorates,
with their associated emergent behavior, provide us
with explicit demonstrations that the underlying microscopic
theory can easily have no measurable consequences whatsoever
at low energies. The nature of the underlying theory is unknowable
until one raises the energy scale sufficiently to escape
protection''.
The notion of \emph{quantum protectorate} was
introduced to unify some generic features of complex physical
systems on different energy scales,  and is a complimentary unifying idea
resembling the symmetry breaking concept in a certain sense. \\
In the search for a "theory of everything"~\cite{pnas}, scientists scrutinize ever-smaller
components of the universe. String theory postulates units so minuscule that researchers
would not have the technology to detect them for decades. R.B. Laughlin~\cite{pnas,laf05,laf08,pines}, argued
that smaller is not necessarily better. He proposes turning our attention instead to
emerging properties of large agglomerations of matter. For instance, chaos theory has been
all the rage of late with its speculations about the "butterfly effect," but understanding how
individual streams of air combine to form a turbulent flow is almost impossible~\cite{kle11}. It may
be easier and more efficient, says Laughlin, to study the turbulent flow. Laws and theories
follow from collective behavior, not the other way around, and if one will try to analyze
things too closely, he may not understand how they work on a macro level. In many cases,
the whole exhibits properties that can not be explained by the behavior of its parts. As
Laughlin points out, mankind use computers and internal combustion engines every day,
but scientists do not totally understand why all of their parts work the way they do.\\
The authors formulate their main thesis: emergent physical
phenomena, which are regulated by higher physical
principles, have a certain property, typical for these
phenomena only. This property is their insensitivity to
microscopic description. For instance,
the crystalline state is the simplest known example of a
quantum protectorate, a stable state of matter whose generic
low-energy properties are determined by a higher organizing
principle and nothing else. There are many other examples~\cite{pnas}.
These quantum protectorates,
with their associated emergent behavior, provide us
with explicit demonstrations that the underlying microscopic
theory can easily have no measurable consequences whatsoever
at low energies. The nature of the underlying theory is unknowable
until one raises the energy scale sufficiently to escape
protection. The existence of two scales, the low-energy and
high-energy scales, relevant to the description of magnetic
phenomena was stressed by the author of this
report in the papers~\cite{kuz02,kuz09} devoted to comparative
analysis  of localized and band models of
quantum theory of magnetism. It was shown there, that
the low-energy spectrum of magnetic excitations in the
magnetically-ordered solid bodies corresponds to a
hydrodynamic pole ($\vec{k}, \omega \rightarrow 0 $) in the generalized
spin susceptibility  $\chi$, which is present in the Heisenberg,
Hubbard, and the combined $s-d$ model. In
the Stoner band model the hydrodynamic pole is
absent, there are no spin waves there. At the same time,
the Stoner single-particle's excitations are absent in the
Heisenberg model's spectrum. The Hubbard model  
with narrow energy bands contains both types
of excitations: the collective spin waves (the low-energy
spectrum) and Stoner single-particle's excitations
(the high-energy spectrum). This is a big advantage
and flexibility of the Hubbard model in comparison
to the Heisenberg model. The latter, nevertheless, is
a very good approximation to the realistic behavior in
the limit $\vec{k}, \omega \rightarrow 0,$
the domain where the hydrodynamic description is
applicable, that is, for long wavelengths and low energies.
The quantum protectorate concept was applied to
the quantum theory of magnetism by the author of this
report in the paper~\cite{kuz02}, where a criterion of applicability of models
of the quantum theory of magnetism  to
description of concrete substances was formulated. The
criterion is based on the analysis of the model's low-energy
and high-energy spectra. 
%
%
%
\section{Conclusions and Discussion}
%
In our interdisciplinary review~\cite{kuz10} we analyzed the applications of the symmetry
principles to quantum and statistical physics in connection with some other branches of
science. The profound and innovative idea of quasiaverages formulated by
N.N.Bogoliubov, gives the so-called macro-objectivation of the degeneracy in domain of
quantum statistical mechanics, quantum field theory and in the quantum physics in
general. We discussed also the complementary unifying ideas of modern physics, namely:
spontaneous symmetry breaking, quantum protectorate and emergence. The interrelation
of the concepts of symmetry breaking, quasiaverages and quantum protectorate was
analyzed in the context of quantum theory and statistical physics.
The main aim of that paper were to demonstrate the connection and interrelation of these
conceptual advances of the many-body physics and to try to show explicitly that those
concepts, though different in details, have a certain common features. Many problems in
the field of statistical physics of complex materials and systems (e.g. the chirality of
molecules) and the foundation of the microscopic theory of magnetism~\cite{kuz02} and
superconductivity~\cite{kuz09} were discussed in relation to these ideas.
It is worth to emphasize once again that the notion of quantum protectorate complements
the concepts of broken symmetry and quasiaverages by making emphasis on the hierarchy
of the energy scales of many-particle systems. In an indirect way these aspects of
hierarchical structure arose already when considering the scale invariance and
spontaneous symmetry breaking in many problems of classical and quantum physics.\\
It was shown also in papers~\cite{kuz10,kuz10a,kuz09}  that the concepts of symmetry breaking perturbations
and quasiaverages play an important role in the theory of irreversible processes as well.
The method of the construction of the nonequilibrium statistical operator~\cite{kuz07} becomes
especially deep and transparent when it is applied in the framework of the Bogoliubov's
quasiaverage concept. For detailed discussion of the Bogoliubov's ideas and methods in the
fields of nonlinear oscillations and nonequilibrium statistical mechanics see Ref.~\cite{kuz07}.
Thus, it was demonstrated in Ref.~\cite{kuz10} that the connection and interrelation of the
conceptual advances of the many-body physics discussed above show that those concepts,
though different in details, have complementary character.\\
To summarize, the ideas of symmetry breaking, quasiaverages, emergence and quantum
protectorate play constructive unifying role in modern theoretical physics. The main
suggestion is that the emphasis of symmetry breaking concept is on the symmetry itself,
whereas the method of quasiaverages emphasizes the degeneracy of a system. The idea of
quantum protectorate reveals the essential difference in the behavior of the complex
many-body systems at the low-energy and high-energy scales. Thus the role of symmetry
(and the breaking of symmetries) in combination with the degeneracy of the system was
reanalyzed and essentially clarified within the framework of the method of quasiaverages.
The complementary notion of quantum protectorate might provide distinctive signatures
and good criteria for a hierarchy of energy scales and the appropriate emergent behavior.
It was demonstrated also that the Bogoliubov's method of quasiaverages plays a
fundamental role in equilibrium and nonequilibrium statistical mechanics and quantum
field theory and is one of the pillars of modern physics.\\ 
We believe that all these concepts
will serve for the future development of physics~\cite{gro05} as useful practical tools. Additional
material and discussion of these problems can be found in recent publications~\cite{ref18,ref19}.
\end{document}